\documentclass[apjl]{emulateapj}

\submitted{Submitted \today}

\shorttitle{Stellar migration and chemical evolution}
\shortauthors{Spitoni et al.}

\begin{document}

\title{The effect of stellar migration on Galactic chemical evolution: 
a heuristic approach}

\author{Emanuele  Spitoni\altaffilmark{1}, Donatella Romano\altaffilmark{2}, Francesca Matteucci \altaffilmark{1,3,4}, and Luca Ciotti \altaffilmark{5}  }

\email{Emanuele Spitoni spitoni@oats.inaf.it}

\altaffiltext{1}{Dipartimento di Fisica, Sezione di Astronomia, Universit\`a di Trieste, via G.B. Tiepolo 11, I-34131, Trieste, Italy  }
\altaffiltext{2}{ I.N.A.F., Osservatorio Astronomico di Bologna, Via Ranzani 1, I-40127 Bologna, Italy}
\altaffiltext{3}{I.N.A.F. Osservatorio
  Astronomico di Trieste, via G.B. Tiepolo 11, I-34131, Trieste,
  Italy}
\altaffiltext{4}{I.N.F.N. Sezione di Trieste, via Valerio 2, 34134 Trieste, Italy}
\altaffiltext{5}{Dipartimento di Fisica e Astronomia, Universit\`a di Bologna, viale Berti-Pichat 6/2, 40127 Bologna, Italy}

\begin{abstract}
 
In the last years, stellar migration in galactic discs has been the
subject of several investigations.  However, its impact on the
chemical evolution of the Milky Way still needs to be fully
quantified. In this paper, we aim at imposing some constraints on the
significance of this phenomenon by considering its influence on the
chemical evolution of the Milky Way thin disc. We do not investigate
the physical mechanisms underlying the migration of stars. Rather, we
introduce a simple, heuristic treatment of stellar migration in a
detailed chemical evolution model for the thin disc of the Milky Way,
which already includes radial gas flows and reproduces several
observational constraints for the solar vicinity and the whole
Galactic disc. When stellar migration is implemented according to the
results of chemo-dynamical simulations by Minchev et. al. (2013) and
finite stellar velocities of 1 km s$^{-1}$ are taken into account, the
high-metallicity tail of the metallicity distribution function of
long-lived thin-disc stars is well reproduced. By exploring the
velocity space, we find that the migrating stars must travel with
velocities in the range 0.5 -2 km s$^{-1}$ to properly reproduce the
high-metallicity tail of the metallicity distribution.  We confirm
previous findings by other authors that the observed spread in the
age-metallicity relation of solar neighbourhood stars can be explained
by the presence of stars which originated at different Galactocentric
distances, and we conclude that the chemical properties of stars currently
observed in the solar vicinity do suggest that stellar migration is
present to some extent.

\end{abstract}


\section{Introduction}

Chemical evolution models are fundamental tools to understand the
formation and evolution of the Milky Way  (e.g. Matteucci
  2001). A good agreement between model predictions and observed
properties of the Galaxy  is usually obtained by assuming
  that the disc is well described by concentric annular regions
  forming by continuous infall of gas without any exchange of matter
  between them (e.g. Matteucci \& Fran\c{c}ois 1989; Chiappini et
  al. 1997; Fran\c{c}ois et al. 2004, among many others).

However, in the presence of gas infall, the Galactic disc cannot be
adequately described by a model with non-interacting zones (see, e.g.,
Spitoni \& Matteucci 2011, and references therein). In particular,
radial motions of gas should be taken into account when the infalling
gas  has a specific angular momentum lower than that of
  the matter moving on circular orbits in the disc, so that mixing
with the gas in the disc induces a net radial inflow.  Lacey \& Fall
(1985) estimated that the gas inflow velocity is approximately 1 km
s$^{-1}$ at 10 kpc. Other internal mechanisms, in principle able to
induce radial gas flows, may also be present in disc galaxies, as a
consequence of asymmetric drift (Smet, Posacki \& Ciotti 2014).  Goetz
\& Koeppen (1992), Portinari \& Chiosi (2000), Spitoni \& Matteucci
(2011), Bilitewski \& Sch\"onrich (2012), Spitoni et al. (2013), Mott
et al. (2013) and Cavichia et al. (2014) present chemical evolution
models with the inclusion of radial gas flows. All these works
generally agree on the magnitude and patterns needed for
  the gas velocities to fit the observational data: the maximum
velocity must not exceed a few km s$^{-1}$.

On the other hand, in all the models cited above the stars are assumed
to remain at their birth radii.  The effect of the interaction of the
stars with the spiral arms on the stellar orbits is discussed by
Sellwood \& Binney (2002), who show that: a) stars on circular orbits
experience larger angular momentum changes, and consequently suffer
more radial migration, than stars on low-angular momentum orbits; b)
stars can migrate to large distances within discs, whilst remaining on
nearly circular orbits. By studying the specific case of the Milky
Way, Sellwood \& Binney (2002) found that old stars formed in the
solar neighbourhood may have been scattered nearly uniformly within
the annular region  between 4 and 12 kpc from the
  Galactic centre --a scenario confirmed qualitatively by subsequent
  studies. For instance, R\v{o}skar et al. (2008, 2012), using
  high-resolution N-body simulations coupled with smoothed particle
  hydrodynamics (SPH), have shown that stars can migrate across large
  Galactocentric distances due to resonant scattering with spiral
  arms, while remaining on almost circular orbits. The spiral
perturbations redistribute angular momentum within the disc and lead
to substantial radial displacements of individual stars, in a manner
that largely preserves the circularity of their orbits.  These models
are able to explain the observed flatness and spread in the
age-metallicity relation of solar neighbourhood stars.
  Recent work by Vera-Ciro et al. (2014) confirm that stellar
migration can be driven by Sellwood \& Binney's (2002) corotation
scattering process.

All the above mentioned dynamical models do not consider the effect of
the Galactic bar;  the role of the bar for radial
  migration is thoroughly analysed in Minchev \& Famaey (2010) and
  Minchev et al. (2011). These authors suggest that radial migration
  results from the nonlinear coupling between the bar and the spiral
  waves and predict a variation of the efficiency of migration with
  time and Galactocentric radius. In a more recent study, Halle et
  al. (2015) confirm that strong radial migration has to be expected
  in the case of a bar-spiral resonance overlap, but also suggest that
  the Galactic outer Lindblad resonance acts as a barrier: stars do
  not migrate from the inner to the outer disc, and vice
  versa\footnote{According to Dehnen (2000), the outer Lindblad
    resonance lies in the vicinity of the Sun.}.

Although the existence of the phenomenon of radial migration seems to
be established beyond any doubt, {\it a quantitative estimate of its
  impact on the chemical evolution of the Galactic disc is much more
  difficult}. Besides the above-mentioned work of Halle et al. (2015),
that challenges any scenario in which significant fractions of stars
move from the inner to the outer disc, and vice versa, some concerns
about the effective role of the bar in stellar migration have been
raised by S\'anchez-Bl\'azquez et al. (2014). These authors present a
comparative study of the stellar metallicity gradients and age
distributions in a sample of nearly face-on spiral galaxies, with and
without bars. The process of radial migration should flatten the
stellar metallicity gradient with time and, therefore, flatter stellar
metallicity gradients should be expected in barred galaxies. However,
 S\'anchez-Bl\'azquez and collaborators do not find any
  difference in the metallicity nor in the age gradients of galaxies
  with or without bars.

The first attempt to analyse the impact that stellar migration could
have on the chemical properties of the Milky Way disc was done by
Fran\c{c}ois \& Matteucci (1993).  By discussing the spread observed
in the age-metallicity relation and in the [$\alpha$/Fe] versus [Fe/H]
relations for solar neighbourhood stars, these authors concluded that
most of the observed spread can be accounted for by the fact that some
of the stars we observe at the present time in the solar vicinity
might have been born in different regions of the Galactic disc, in
particular in more inner ones.  Their paper was based on observational
features of a possible radial migration presented in pioneering works
by Grenon (1972, 1989). Grenon identified an old population of super
metal-rich stars, that are currently located in the solar vicinity,
but have kinematics and abundance properties indicative of an origin
in the inner Galactic disc.

Despite the recent advances in the field of galaxy formation and
evolution, there are currently no self-consistent simulations
containing the level of chemical implementation required for making
detailed predictions for the number of ongoing and planned Milky Way
observational campaigns. Even in high-resolution N-body experiments,
one particle represents $\approx 10^{4-5}$ solar masses. Hence, a
number of approximations are necessary to compute the chemical
enrichment self-consistently inside a dynamical simulation.

Sch\"onrich \& Binney (2009a,b) included in a chemical model the
radial migration of stars and gas through the disc of the Galaxy by
means of a probabilistic approach constrained by Milky Way observables
and found results in agreement with R\v{o}skar et al. (2008). They
found also that most observational properties and peculiarities of the
thick disc could be explained by radial migration. Recently, Minchev
et al. (2013, 2014) have presented a new approach for combining
chemistry and dynamics that overcomes the technical problems
originated by fully self-consistent simulations, and avoids the star
formation and chemical enrichment problems encountered in previous
simulations. They assume that each particle is one star and implement
the star formation history and chemical enrichment deriving from a
classic chemical evolution model for the thin disc into the simulated
Galactic disc. Kubryk et al. (2013, 2014) also studied radial
migration and chemical evolution for a general bar-dominated disc
galaxy, by analysing the results of a fully self-consistent,
high-resolution N-body+SPH simulation. Stars undergo substantial
radial migration at all times, caused first by transient spiral arms
and later by the bar. The authors stress that, despite the important
amount of radial migration occurring in their model, its impact on the
chemical properties is limited.

Here,  we adopt a phenomenological approach to study the
  effects of different prescriptions for the stellar migration on the
best known chemical evolution observables. In practice, we implement
stellar migration in a very simple way in a detailed chemical
evolution model for the Galactic thin disc that already
includes radial gas flows. We first focus on the metallicity
distribution function of long-lived stars in the solar
neighbourhood. We then examine the effects of the migration on other
observables, such as the age-metallicity relation of stars in the
solar vicinity and the stellar density profile across the disc.

The paper is organised as follows: in Section~2 we describe our
reference chemical evolution model and the way in which the radial gas
flows and the stellar migration have been implemented. In Section~3 we
report the model results without stellar migration, whereas model
results with stellar migration are presented in Section~4. Finally, we
draw our main conclusions in Section~5.

\begin{table*}[t]
\caption{Chemical evolution models}
\scriptsize 
\label{cem}
\begin{center}
\begin{tabular}{c|cccccccc}
  \hline

 Models &$\tau_D$&$\nu$& Radial gas inflow & Stellar migration& Stellar velocities\\
&[Gyr]&[Gyr$^{-1}$]&[km s$^{-1}$] &&[km s$^{-1}$]\\  
\hline

REF1& 1.033 R[kpc]-1.27&variable &1& /&/\\
&&  (see Figure \ref{eff}) &&\\
\hline

REF2& 1.033 R[kpc]-1.27& 1 &1& /&/\\
\hline
`Minchev case'& 1.033 R[kpc]-1.27&variable &1&10 \% from 4 kpc &0.5, 1, 2\\
& &(see Figure \ref{eff})   &&20 \% from 6 kpc &\\
& &   &&60 \% from 8 kpc &\\
\hline

 \hline
\end{tabular}
\end{center}
\end{table*}

\section{The chemical evolution model}

 Our reference model for the Galactic thin disc is that
  presented in Spitoni \& Matteucci (2011). The model assumes that
the disc grows via smooth accretion of gas of primordial chemical
composition. The infall law is:
\begin{equation}\label{infall} 
A(r,t)= a(r)e^{-\frac{t}{\tau_D}},
\end{equation} 
 where $\tau_D$, the $e$-folding timescale, is a free
  parameter of the model and the coefficient $a(r)$ is set by the
  requirement that the current total mass surface density profile of
  the Galactic thin disc is reproduced. In order to mimic an
  inside-out formation of the disc (Larson 1976), the timescale for
  mass accretion is assumed to increase with the Galactic radius
  following a linear relation (see Chiappini et al. 2001, and
  references therein):
\begin{equation}
\tau_{D}(r)=1.033 r (\mbox{kpc}) - 1.27 \,\, \mbox{Gyr};
\label{t1}
\end{equation}
this relation holds for Galactocentric distances between 4 and
16~kpc. The region within 4~kpc is not considered in this
paper. The infall law adopted in this work is clearly a
  rough approximation of the true mass assembly history of the disc,
still necessary since we do not rely on a full cosmological
setting. However, it has been shown (Colavitti et al. 2008)
that the results of a model assuming such an
  exponentially-decaying infall law are fully compatible with those
  obtained with cosmological accretion laws.

As for the IMF, we use that of Scalo (1986), constant in time and
space. We indicate with $\tau_{m}$ the evolutionary lifetime of the
stars with initial mass {\it m} (Maeder \&Meynet 1989).  The Type Ia
supernova (SN) rate is computed following Greggio \& Renzini (1983)
and Matteucci \& Greggio (1986).  The stellar yields are taken from
Woosley \& Weaver (1995) for core-collapse SNe, van den Hoek \&
Groenewegen (1997) for low- and intermediate-mass stars, and Iwamoto
et al. (1999) for Type Ia SNe. We use a star formation (SF) rate
proportional to the gas surface density:
 
\begin{equation} 
\psi(r,t) \propto \nu \sigma_{gas}^k(r,t) 
\end{equation} 
where $\nu$ is the efficiency of the SF process and
  $\sigma_{gas}(r,t)$ is the gas surface density at a given position
  and time. The exponent $k$ is fixed to 1.4 (see Kennicutt 1998).

\begin{figure}[!h]
	  \centering   
    \includegraphics[scale=0.42]{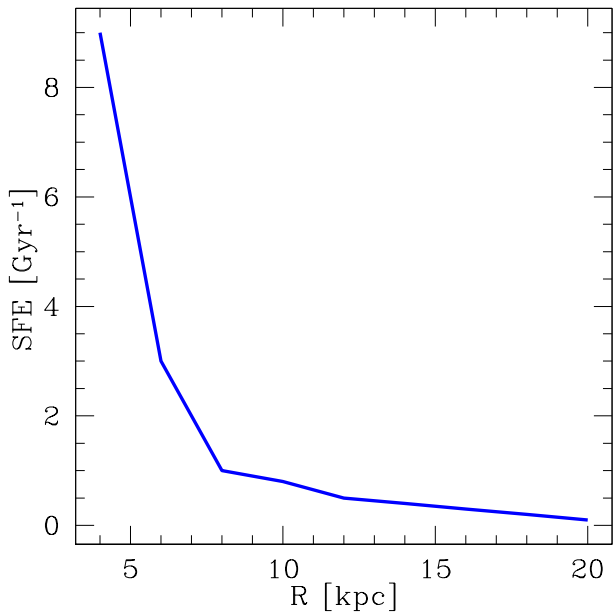}  
    \caption{Adopted star formation efficiency as a function of Galactocentric radius   for our models (except model~REF2, that assumes a constant star formation efficiency).}
		\label{eff}
\end{figure}

Motivated by the theory of SF induced by spiral density waves in
galactic discs (Wyse \& Silk 1989), we consider a variable star
formation efficiency (SFE) as a function of the Galactocentric
distance, $\nu \propto 1/r$  (model REF1, Table~1).
This is different from Spitoni \& Matteucci (2011), who use a constant
value for the SFE  (model REF2, Table~1).  The reason
for this choice is related to the fact that with a constant SFE the
predicted abundance ratio patterns (e.g. [O/Fe] vs [Fe/H]) do not
change much from radius to radius, thus making the hypothesis of
stellar migration as a solution for the observed spread rather
unlikely. Figure ~\ref{eff} shows the SFE as a function of the
Galactocentric distance used in this work.  It is worth
  emphasising that a larger SFE in the inner disc is assumed also by
  Kubryk et al. (2014) and, probably, by Minchev et al. (2013); in the
  latter paper the authors do not state it explicitly, but an
  inspection of their results leads us to believe that this is the
  case. It is the combination of radial migration with a variable SFE
across the disc that allows to reproduce the observed abundance
spreads.

\subsection{The implementation of the gas radial flows}

In our reference model a constant radial inflow of gas
  with velocity of -1 km s$^{-1}$ is implemented following the
  prescriptions of Spitoni \& Matteucci (2011).

We define the $k$-th shell in terms of the Galactocentric radius
$r_k$, its inner and outer edge being labeled as $r_{k-\frac{1}{2}}$
and $r_{k+\frac{1}{2}}$.  We take the inner edge of the $k$-shell,
$r_{k-\frac{1}{2}}$, at the midpoint between the characteristic radii
of the shells $k$ and $k-1$, and similarly for the outer edge
$r_{k+\frac{1}{2}}$. The flow velocities $v_{k+\frac{1}{2}}$ are
assumed to be positive outward and negative inward.

Radial inflows  with a flux $F(r)$ contribute to alter the gas surface density $\sigma_{gk}$  in the $k$-th  shell according to
\begin{equation}
\label{dsigmarf1}
\left[ \frac{d \sigma_{g k}}{d t} \right]_{rf} = 
   - \frac{ F(r_{k+\frac{1}{2}}) - F(r_{k-\frac{1}{2}}) }{\pi \left( r^2_{k+\frac{1}{2}} - r^2_{k-\frac{1}{2}} \right) }
  ,
\end{equation}
where the gas flow at $r_{k+\frac{1}{2}}$ can be written as
\begin{small}
\begin{equation}
\label{flux3}
F(r_{k+\frac{1}{2}}) = -2 \pi \, r_{k+\frac{1}{2}} \,  v_{k+\frac{1}{2}} \, \sigma_{g (k+1)}.
\end{equation}
\end{small}
We assume that there are no flows from the outer parts of the disc where there is no SF. In our implementation of the radial inflow of gas, only the gas in the Galactic disc at $r <$ 20 kpc can move inward by radial inflow.

\subsection{Including stellar migration in our chemical evolution model}

As discussed in the Introduction,  we implement stellar
  migration in our chemical evolution code using a phenomenological
  approach, without a specific description of the complex dynamical
  mechanisms affecting stellar orbits in galactic discs. In practice,
we simply move stars formed at a given radius, with a certain age and
a certain metal content, to the solar vicinity, 
  according to some simple rule. Following the work of Kordopatis et
al. (2013), we assume as a representative value for the velocity of
the stars 1 km s$^{-1}$, that is $\simeq$ 1~kpc
Gyr$^{-1}$. To explore the velocity parameter space we
  also run models with stellar velocities fixed to 0.5 and 2 km
s$^{-1}$.

We stress that by taking into account a velocity of 1~km s$^{-1}$,
only stars with stellar  lifetimes $\tau_m \ge$~1~Gyr,
  corresponding to initial masses $m \le$~2~M$_{\odot}$ (see figure~3
of Romano et al. 2005, and references therein), can travel for
distances larger than 1~kpc. We can thus safely assume that most of
the metals produced by a stellar generation are ejected well within
1~kpc from  their progenitors' birthplace. Therefore,
the metal enrichment of the ISM in the solar vicinity is 
  practically unaffected by the stars born in the inner regions, that
  are thus only `passive tracers' of the process of
migration.  We notice that, as estimated by Lacey \& Fall
  (1985) with simple calculations, the magnitude of the radial gas
  flows velocity is roughly 1 km s$^{-1}$ at 10 kpc. Hence, stellar
  migration and radial gas inflows are characterised by the same time
  scales, and this is the reason why one has to consider both
  processes for a complete and consistent study.

First, we want to test the effect of  assuming a
  physically plausible stellar migration on our chemical evolution
 model predictions. To this aim, we use the results of
  chemo-dynamical simulations by Minchev et al. (2013) to obtain
  general recipes to be used in our pure chemical evolution code. We
  choose this study since the chemical evolution model adopted by the
  authors is very similar to ours, which allows a test of the validity
  of our approach. Indeed, as we will show in Section 4, if we assume
  that 10\% and 20\% of the stars born at 4 and 6 kpc, respectively,
  migrate towards 8 kpc, while 60\% of those born at 8 kpc leave the
  solar neighbourhood (percentages are drawn from Minchev et al.'s
  dynamical simulations), the metallicity distribution function we
  predict for long-lived solar vicinity stars turns out to be very
  similar to that obtained by Minchev and collaborators.

\begin{figure}[!h]
	  \centering   
    \includegraphics[scale=0.42]{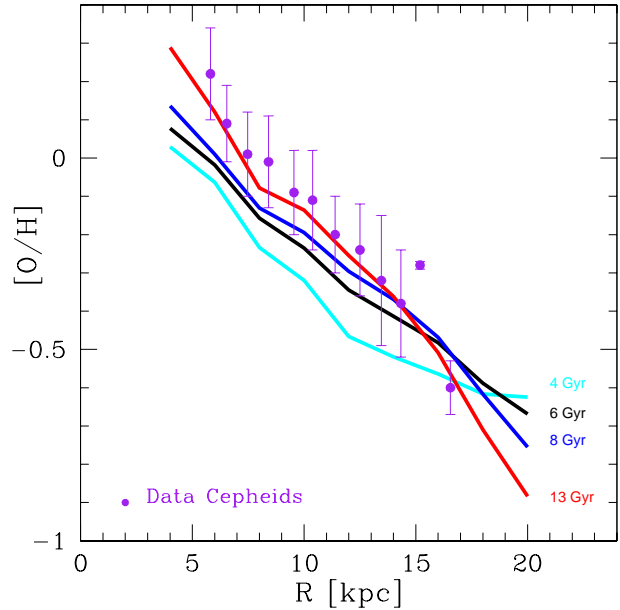} 
    \caption{Evolution in time of the oxygen abundance gradient along the Galactic disc for our reference model in the presence of radial gas flows and variable star formation efficiency (model REF1); shown are the predictions of the model at 4, 6, 8 and 13~Gyr.  The model results are compared to data from Cepheids (Luck \& Lambert 2011), that track the present-day gradient.}
		\label{evol}
\end{figure}

Then, we also test some extreme cases, that are not supported by dynamical studies, with the aim of exploring the effects that a huge migration of stars would have on the chemical evolution of the solar neighbourhood (see Section~4).

\begin{figure}[!h]
	  \centering   
    \includegraphics[scale=0.42]{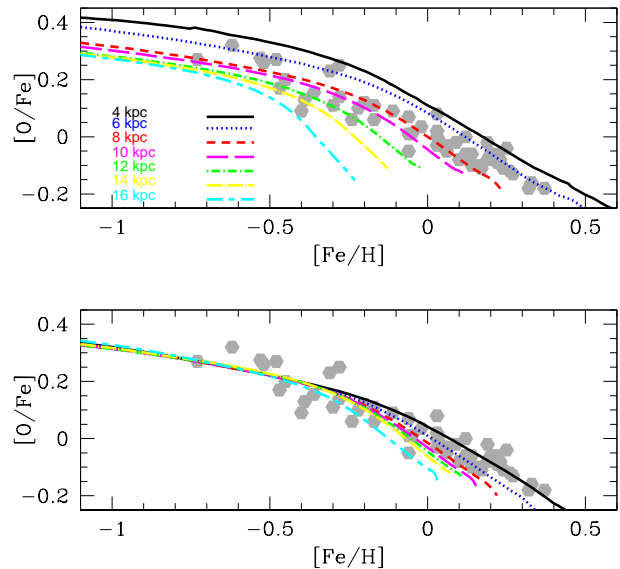} 
    \caption{  [O/Fe] vs [Fe/H] relations (lines)
        predicted by our models REF1 {\it (upper panel)} and REF2 {\it
          (lower panel)} at different Galactocentric distances, as
        indicated in the bottom left legend in the top panel. The two
        models include radial gas flows and differ only in the assumed
        SFE (see text). The data (symbols in both panels) refer to
        solar neighbourhood thin-disc stars and are taken from Bensby
        et al. (2005).}
		\label{OFE_ref}
\end{figure} 

\begin{figure}[!h]
	  \centering   
    \includegraphics[scale=0.42]{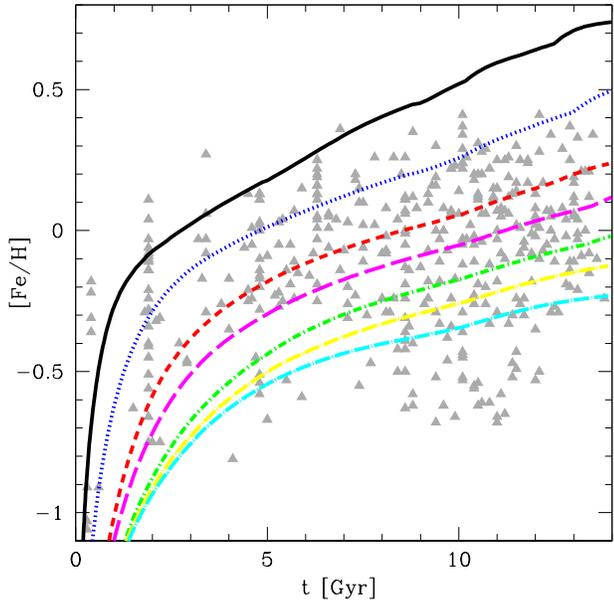} 
    \caption{Temporal evolution of [Fe/H] for our best model REF1 with
      radial gas flows and variable SFE, at different Galactocentric
      distances (curves; see Figure \ref{OFE_ref}, upper panel, for
      the legend). Also show are data by Bensby et al. (2014) for
      solar neighbourhood stars (symbols).}
		\label{Fe_t}
\end{figure}

\begin{figure}[!h]
	  \centering   
  \includegraphics[scale=0.42]{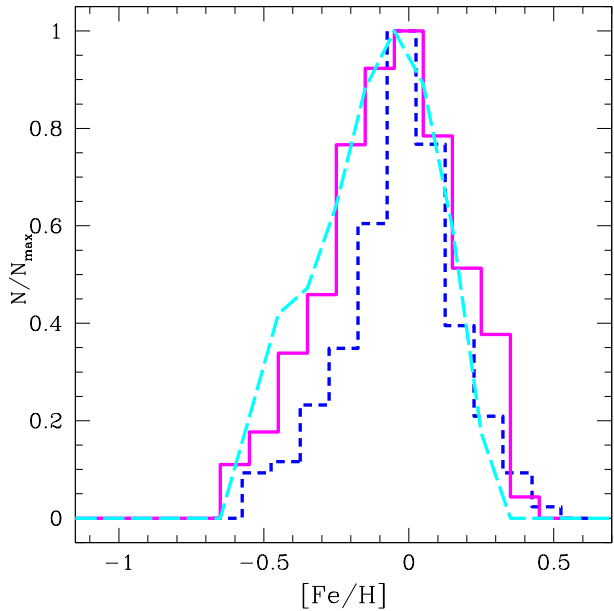} 
   \caption{The G-dwarf metallicity distribution predicted by
     model~REF1 (dashed line), that does not include stellar migration
     but includes a constant radial gas inflow of 1 km s$^{-1}$, is
     compared to the observed distributions of thin-disc stars by
     Adibekyan et al. (2013; solid magenta histogram) and Fuhrmann
     (2011; short-dashed blue histogram).}
		\label{G_ref_f}
\end{figure}

\begin{figure}[!h]
	  \centering   
    \includegraphics[scale=0.42]{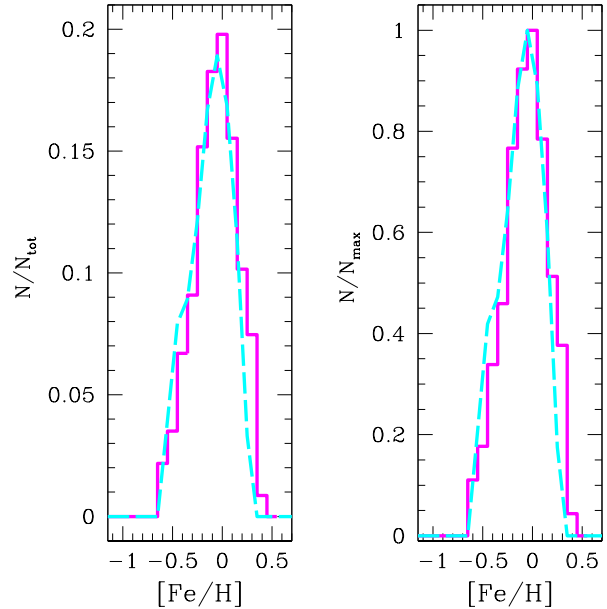} 
    \caption{ The G-dwarf metallicity distribution predicted by model
      REF1 without stellar migration (dashed line) is compared to the
      observed one by Adibekyan et al. (2013; solid
      histogram).  The distributions are normalised to
        either the total or the maximum number of stars in each
        distribution ({\it left} and {\it right panels,}
        respectively).}
		\label{G_ref}
\end{figure}

\section{Model results without stellar migration}

 In this section, we present the results of our chemical
  evolution model including radial gas flows, either with variable
  (model REF1) or constant SFE (model REF2).

In Table \ref{cem}, the principal characteristics of the 
  models are reported: the time scale of thin-disc formation,
  $\tau_D$, is listed in column 2; the adopted SFEs in column 3; the
  velocity of the radial gas inflow in column 4. For the model
  including stellar migration (`Minchev case' model), columns 5 and 6
  specify the adopted prescriptions about the fraction of migrating
  stars and their velocities.

In Figure \ref{evol} it is shown that  model REF1 (red
  curve, labelled 13 Gyr) perfectly fits the present-day oxygen
  abundance gradient measured in Cepheids. The data (from Luck \&
  Lambert 2011) have been divided into bins 1-kpc wide; for each bin
  we compute the average value and the relative standard deviation. In
  Figure \ref{evol} we also show the evolution of the gradient,
  computed at 4, 6, 8 and 13 Gyr. It is worth noting that the
  combination of a variable SFE --with higher values in the inner disc
  regions-- with a constant radial gas flow leads to a steepening of
the gradient in time.

In the upper panel of Figure \ref{OFE_ref} we show the [O/Fe] vs
[Fe/H] relations obtained  with model~REF1 at different
  Galactocentric distances, spanning the range between 4 and 16~kpc.
  The theoretical predictions are very similar to the ones reported by
  Minchev et al. (2013; see their figure~4, bottom left panel) and
are mainly driven by the adoption of a SFE that varies with the
Galactocentric distance in the model. For reference, we also plot
 data for solar neighbourhood  thin-disc
  stars by Bensby et al. (2005).  In the lower panel of Figure
\ref{OFE_ref} we show the effects of  assuming a constant
  SFE (model REF2) on the [O/Fe] vs [Fe/H] diagram. It is immediately
  seen that the observed spread in the data can not be longer
  reproduced by the chemical evolution model. We conclude that the
hypothesis of stellar migration as a solution for the observed spread
requires necessarily a variable SFE.   This is even more
  evident from Figure \ref{Fe_t}, where we show the age-metallicity
  relation predicted by model~REF1 for different Galactocentric
  distances, compared to the data for thin-disc stars in the solar
  vicinity from Bensby et al. (2014). Other authors already claimed
(e.g. Fran\c{c}ois \& Matteucci 1993; Sch\"onrich \& Binney 2009a)
that the observed spread in the data can be explained with the
migration of stars both from the inner and the outer regions, and we
confirm their findings. Also, we remark that, while the migration
produces a spread of up to 1 dex in [Fe/H] at a given age, a much
lower spread is expected in the [O/Fe] ratios at a given metallicity
(cfr. Figures.~\ref{OFE_ref} and \ref{Fe_t}), in agreement with what
is suggested from the observations (see Edvardsson et al. 1993, who
first discussed the conundrum of a large spread in the age-metallicity
relation apparently contrasting with the tightness of the [element/Fe]
versus [Fe/H] relations for solar neighbourhood stars).

\begin{figure}[!h]
	  \centering   
    \includegraphics[scale=0.42]{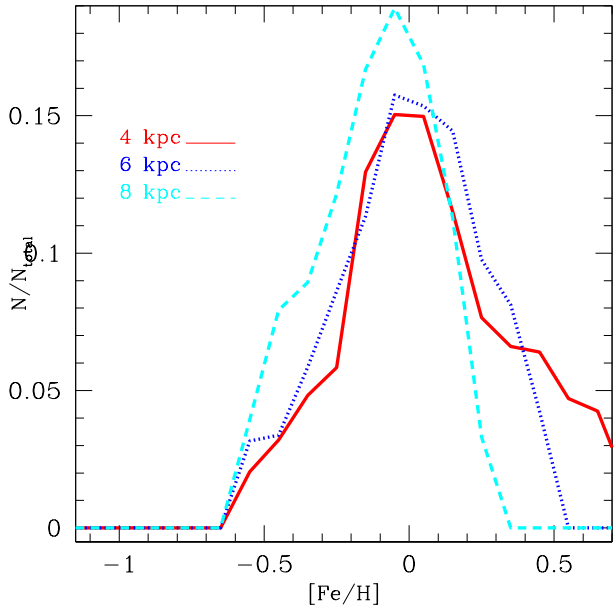} 
    \caption{The G-dwarf  metallicity distributions predicted by model~REF1 at 4, 6, and 8 kpc, assuming radial gas flows and no stellar migration.}
		\label{468}
\end{figure} 

In Figure \ref{G_ref_f} we  compare the metallicity
  distribution function of long-lived, thin-disc stars predicted by
  model~REF1 (dashed line) to the observed ones from Adibekyan et
  al. (2013; solid histogram) and Fuhrmann (2011; dashed
  histogram). The sample of Adibekyan et al. (2013), comprising stars
  with Galactocentric distances between 7.9 and 8.1 kpc and with
  heights above the Galactic plane $|Z|$ smaller than 0.1 kpc, is not
  a pure thin-disc sample. However, thin-disc stars can be separated
  from the rest of the sample based on a chemical criterion (the
  [$\alpha$/Fe] vs [Fe/H] plot; see Adibekyan et al. 2013). The
  distributions of Adibekyan et al. (2013) and Fuhrmann (2011), while
  agreeing on the position of the peak, significantly differ in
  shape. This could be due to sample selection effects: while
  Fuhrmann's survey is limited to objects with {\it Hipparcos} data
  within 25~pc from the Sun, Adibekyan et al. (2013) rely on
  high-resolution analysis of a kinematically unbiased sample of more
  than 1000 FGK stars located up to 600~pc --but mostly within 50~pc--
  from the Sun. Interestingly, the APOGEE distribution, extending to
  significantly larger distances and sampling nearly 10 times more
  stars, compares well to that from Adibekyan et al. (2013; see Anders
  et al. 2014). Therefore, in the following we will use the Adibekyan
  et al. (2013) distribution for comparison with the model
  predictions. We also recall that our  Galactic
  chemical evolution model is a `sequential' one, and that we
consider as thin-disc stars only those with [Fe/H] $\geq -$0.6 dex.
We see that the  metallicity distribution function of
Adibekyan et al. (2013) is fairly well reproduced by our model
 with radial gas inflow and no stellar
  migration. However, we note that the model underestimates the
high-metallicity tail of  observed distribution. In
  Figure \ref{G_ref_f}, the theoretical and observed metallicity
  distribution functions are normalised to the corresponding maximum
  number of stars for each distribution. If we normalise the
  distributions to the total number of stars (see Figure \ref{G_ref},
  {\it left panel}), a good agreement is still found between model
  predictions and observations.

\begin{figure}[!h]
  \centering \includegraphics[scale=0.42]{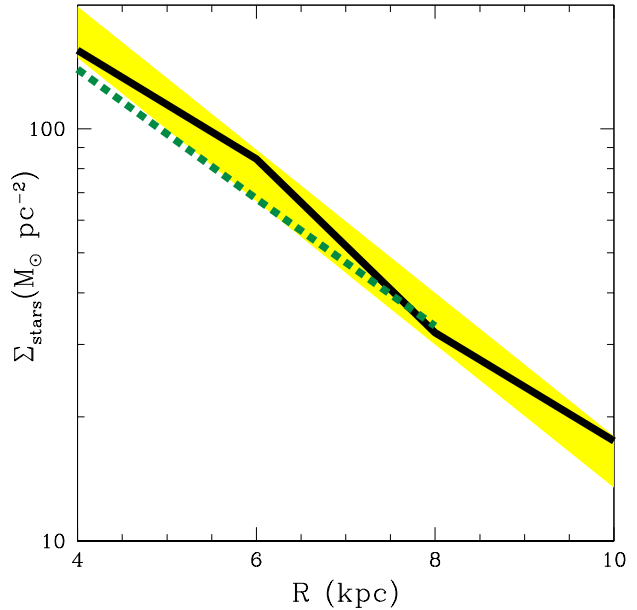} 
\caption{The stellar surface density profile predicted by the chemical
  evolution model REF1 is shown as a black solid line.  The
   green dashed line represents a model in
    which stellar migration is included according to the results of
    chemo-dynamical simulations by Minchev et al. (2013; model labeled
    `Minchev case' in Table~1): 10\% (20\%) of the stars born at 4 kpc
    (6~kpc) move toward 8 kpc, whereas 60\% of the stars born at 8 kpc
    leave the solar neighbourhood.  The yellow stripe} indicates the
  area spanned by the observational data (see text).  \label{sigma}
\end{figure} 

In Figure \ref{468} we show the G-dwarf metallicity distributions
obtained by  model~REF1 at 4, 6, 8 kpc distance from the
  Galactic centre.  As expected, the relative number of
  high-metallicity stars increases going towards the Galactic center,
due to the faster evolution and more efficient star formation
 in the inner regions. These predictions can be compared
  to data from large spectroscopic surveys, such as APOGEE and
  Gaia-ESO, that are mapping increasingly larger volumes of the Milky
  Way disc (e.g., 5~$< R < $~11~kpc, 0~$< |Z| <$~2~kpc, Nidever et
  al. 2014 --APOGEE; 4~$< R <$~12~kpc, 0~$< |Z| <$~3.5~kpc, Mikolaitis
  et al. 2014 --Gaia-ESO Survey).

Finally, in Figure \ref{sigma}, the current stellar surface density
profile predicted by model~REF1 is reported as a black solid line. To
compare our model predictions with the relevant data 
  (yellow stripe in the plot), we consider a local stellar density of
35$\pm$5 M$_{\odot}$ pc$^{-2}$ (Gilmore, Wyse \& Kuijken 1989) and
assume that the stellar profile decreases exponentially outward, with
a characteristic length scale of 2.5~kpc (Sackett 1997).

We stress that all the abundance ratios presented in this section are normalized to our theoretical solar abundances, namely the abundances predicted in the ISM at 8~kpc 4.5~Gyr ago.  In Section~4 we will use the same normalization.

\begin{figure}
	  \centering \includegraphics[scale=0.42]{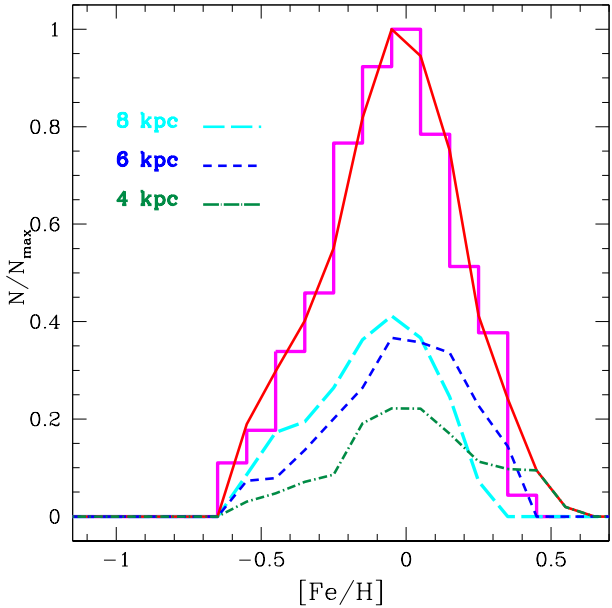} 
 \caption
{  Red solid line: theoretical G-dwarf metallicity
  distribution obtained with the  `Minchev case' model,
    that assumes radial gas flows and stellar migration according to
    the following recipe: 10\% and 20\% of the stars born at 4 and 6
    kpc, respectively, move toward 8 kpc, whereas 60\% of the stars
    born in situ leave the solar neighbourhood; the stars are assigned
    a velocity of 1 km s$^{-1}$. The contributions of the stars born
    at different Galactocentric radii are also shown (legend on top
    left corner).  The data (magenta histogram) are from Adibekyan et
    al. (2013).} \label{comp}
\end{figure}

\section{Model results in the presence of stellar migration}

We now move to illustrate the results of the models that
  include radial stellar migration. First, we set some simple rules to
  make stars travel across the disc, somehow mimicking the results of
  more complex, self-consistent chemo-dynamical models (Minchev et
  al. 2013): we assume that 10\% of the stars born at 4 kpc and 20\%
  of those born at 6 kpc migrate toward 8 kpc, while 60\% of those
  born in the solar neighbourhood migrate  (both outwards and inwards);
  the migrating stars have a velocity of 1~km s$^{-1}$. We refer to
  this model as `Minchev case' model (see Table~\ref{cem}). The
  metallicity distribution function of long-lived solar neighbourhood
  stars obtained with this model is reported in Figure \ref{comp} (red
  solid line). It is clear that we are now able to explain not only
  the peak and left wing of the observed distribution (magenta
  histogram; Adibekyan et al. 2013), but also its high-metallicity
  tail. We also show in Figure \ref{comp} the  separate
  contributions of stars born at 4 and 6 kpc that end up in the solar
  neighbourhood (green dot-dashed and blue dashed lines,
  respectively), as well as that of stars born in situ (light blue
  dashed line). By comparing this figure with figure 3 (middle panel)
of Minchev et al. (2013), we note that our model is entirely
consistent with their results in terms of the G-dwarf metallicity
distribution.  The stellar surface density profile that
  is obtained with the above prescriptions for stellar migration is
  presented in Figure~\ref{sigma} (green dashed line). The observed
  profile (yellow stripe) is reproduced and the model leads to a
stellar surface density of 33 M$_{\odot}$ pc$^{-2}$ in the solar
neighbourhood,  in agreement with the observed value.

\begin{figure}
	  \centering \includegraphics[scale=0.42]{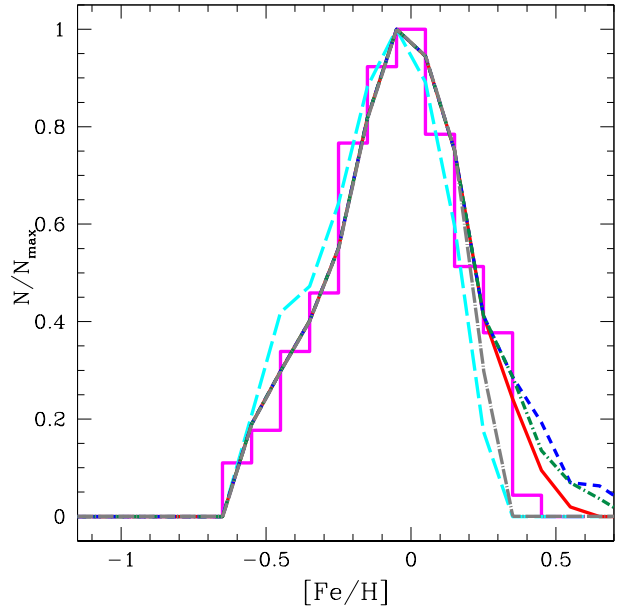} 
 \caption
{ Effects of different prescriptions about the velocity
    of the migrating stars on the metallicity distribution of solar
    neighbourhood long-lived stars predicted with the `Minchev case'
    model. The short-dashed blue line represents the model results
    when the finite velocity of the stars is not taken into
    account. The solid red and short-dashed-dotted green lines refer
    to adopted velocities of 1 and 2 km s$^{-1}$, respectively, while
    the long-dashed-dotted grey line stands for a case in which the
    stellar velocity is set to 0.5 km s$^{-1}$.  With the long-dashed
  cyan line we also show our reference model (REF1) without stellar
  migration. The data (magenta histogram) are for thin-disc stars from
  Adibekyan et al. (2013).}  \label{G_R}
\end{figure} 

In Figure \ref{G_R},  we explore the effects of different
  assumptions about the velocities of the migrating stars on the
  predicted G-dwarf metallicity distribution (the stars migrate
  according to the scheme discussed in the previous paragraph). We
  either do not take the stellar velocity into account, or analyse
  cases in which stellar velocities of 0.5, 1 and 2 km s$^{-1}$ are
  considered. The observed G-dwarf metallicity distribution is best
  reproduced with a velocity of 1 km s$^{-1}$. Adopting a smaller
  velocity (e.g. 0.5 km s$^{-1}$) has the consequence that many more
  stars among the ones born more recently and at larger distances from
  the solar neighbourhood have not sufficient time to reach the solar
  vicinity, compared to models which assume larger stellar
  velocities. As a result, the right tail of the Adibekyan et
al. (2013) distribution is no more reproduced, and the model tends to
the case without stellar migration.

Kobayashi \& Nakasato (2011) presented chemo-dynamical simulations of
a Milky-Way-type galaxy using a self-consistent hydrodynamical code
that includes supernova feedback and chemical enrichment, and predicts
the spatial distribution of elements from oxygen to zinc. With their
model they predict that only 30\% of the stars which reside in the
solar neighbourhood has originated from different regions of the
Galactic disc (C. Kobayashi, private communication). In Minchev et
al. (2013) study, approximately 60\% of the stars 
  currently found in the solar neighbourhood would have formed at 4
  and 6 kpc. Therefore, so far, it is not possible to reach firm
conclusions about the effective amount of stellar migration in the
Galactic disc. It is interesting to note, as already pointed out by
  previous authors, that stellar migration brings to the solar
neighbourhood a few old, metal-rich stars, thus fulfilling the
requirements implied by observations of Grenon (1972,
1989). However, it is also important to stress that it
  does not lead to substantial changes in the theoretical metallicity
  distribution function of long-living stars, making a quantification
  of the extent of migration based on this observable a hard task.

Another important issue is that the migrated stars could end up to populate the thick disc in the solar
neighbourhood, as indicated  by Minchev et al. (2013): in this case, the G-dwarf metallicity distribution would  suffer an even smaller effect by migrating stars.
\begin{figure}[!h]
	  \centering \includegraphics[scale=0.73]{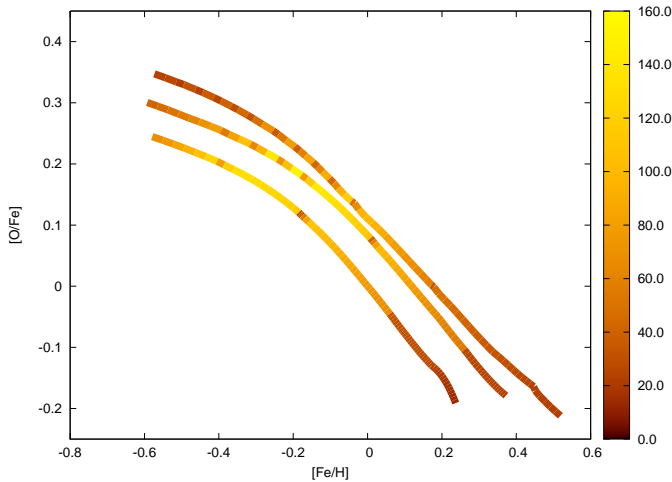} 
\caption{[O/Fe] vs [Fe/H] in the solar neighbourhood for the `Minchev
  case' model, where a velocity of 1 km s$^{-1}$ is considered
  for the migrating stars. The lines are color-coded
    according to the number of stars (in unit of 10$^{6}$).  The upper
  curve refers to stars born at 4 kpc, the middle one to stars born at
  6 kpc, the lower curve represents stars born in
  situ.}  \label{gnuplotofe}
\end{figure}

\begin{figure}[!h]
	  \centering   
    \includegraphics[scale=0.73]{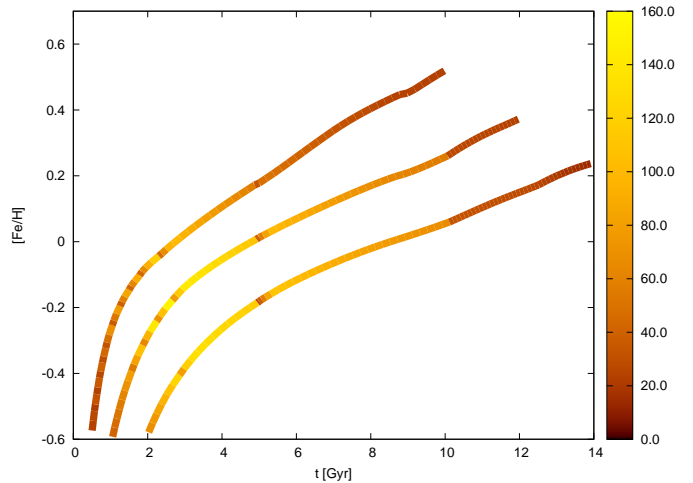} 
    \caption{[Fe/H] vs time in the solar neighbourhood for the
      `Minchev case' model, where a velocity of 1 km s$^{-1}$ is
      considered  for the migrating stars. The lines are
        color-coded according to the number of stars (in unit of
        10$^{6}$). The upper curve refers to stars born at 4 kpc, the
      middle one to stars born at 6 kpc, the lower curve represents
      stars born in situ.}
		\label{gnuplotfe}
\end{figure}

 We compute the numbers of stars in the solar
  neighbourhood that are expected to have a given [O/Fe] at a given
  [Fe/H], as well as the numbers of stars with a certain [Fe/H] at a
  given age, for the `Minchev case' model taking into account a
  stellar velocity of 1 km s$^{-1}$ (see Figures \ref{gnuplotofe}
and \ref{gnuplotfe}, respectively). In Figure \ref{gnuplotfe} it is
clearly shown the effect of  setting the velocity of the
  stars to a fixed value. We are considering velocities fixed at 1 km
  s$^{-1}$ $\approx$ 1 kpc Gyr$^{-1}$ and, thus, the curve relative to
  the stars coming from 4 kpc ends at 10 Gyr: all stars born at 4 kpc
  at times larger than 10 Gyr from the beginning of the simulation do
  not have enough time to reach the solar neighbourhood.  The same
explanation is valid for the 6 kpc line result, i.e. for times larger
than 12 Gyr no stars formed at 6~kpc can reach the solar
  neighbourhood.

\begin{figure}
\centering\includegraphics[scale=0.42]{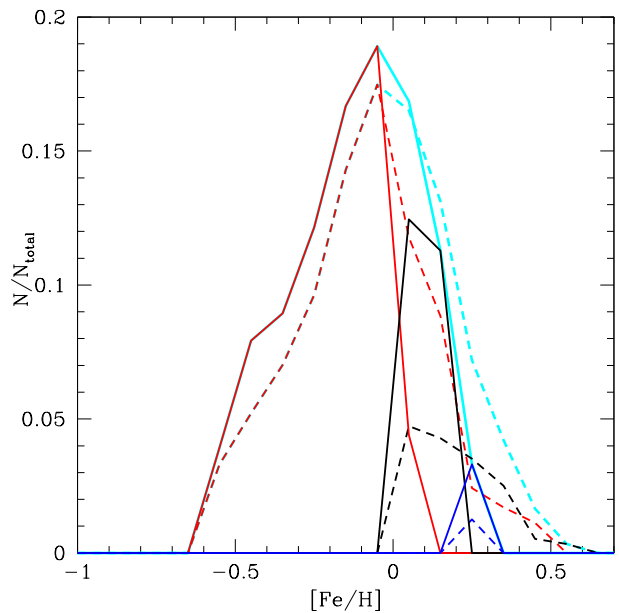}
 \caption{  G-dwarf metallicity distributions at 8 kpc
     obtained by our models for young stars (ages $<$ 1 Gyr; blue
     lines), intermediate-age stars (1 Gyr $<$ ages $<$ 5 Gyr; black
     lines) and old stars (ages $>$ 5 Gyr; red lines). The cyan lines
     show the sum of all age distributions. The solid lines refer to
     model~REF1, without stellar migration but including radial gas
     flows. The dashed lines refer to the `Minchev case' model with
     velocity of the migrating stars fixed at 1 km s$^{-1}$.  Each
     distribution is normalised to the total number
     (young+intermediate+old) of stars.} \label{age}
\end{figure} 

\begin{figure}
	  \centering \includegraphics[scale=0.42]{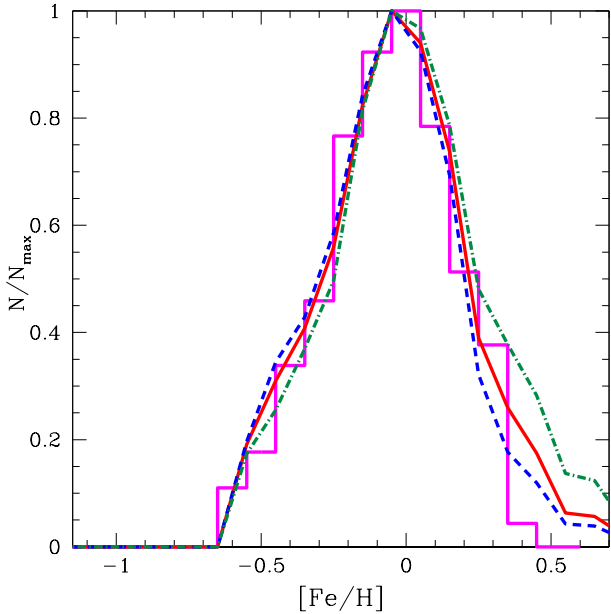} 
 \caption
{  G-dwarf metallicity distributions predicted by our
    `extreme' models where stars born in situ at 8 kpc are not allowed
    to migrate. The dashed blue line refers to the predictions of case
    1) migration (10\% and 20\% of the stars born at 4 and 6 kpc,
    respectively, end up at 8 kpc). Case 2) migration (20\% and 40\%
    of the stars born at 4 and 6 kpc, respectively, end up at 8 kpc)
    is represented by the solid red line and case 3) migration (all
    stars born at 4 and 6 kpc end up at 8 kpc) is represented by the
    green dashed-dotted line.  The observed distribution of thin-disc
    stars (magenta histogram) is taken from Adibekyan et
  al. (2013).}  \label{ex_gd}
\end{figure} 

 In Figure~\ref{age} we show the metallicity
  distributions of long-lived stars in the solar vicinity predicted by
  our models with (dashed lines; `Minchev case' model, velocity of the
  migrating stars set to 1 km s$^{-1}$) and without stellar migration
  (solid lines; model~REF1). The total distributions (cyan lines) are
  split in different age bins: we consider young stars (ages $<$ 1
  Gyr; blue lines), intermediate-age stars (1 Gyr $<$ ages $<$ 5 Gyr;
  black lines) and old stars (ages $>$ 5 Gyr; red lines). Each
  distribution is normalised to the total number
  (young+intermediate+old) of stars. The finite velocity of the stars
  is taken into account in order to establish how many inner-disc
  stars end up in the solar neighbourhood.

In our reference model without stellar migration 
  (model~REF1, with variable SFE and radial gas inflow included with a
  velocity of 1 km s$^{-1}$), 3.30\% of the stars present ages less
than 1 Gyr, 23.74\% have ages between 1 and 5 Gyr, whereas the oldest
stars with ages larger than 5 Gyr are 72.96\% of the total.
Concerning the model with stellar migration (`Minchev case' model),
1.25\% of the stars have ages less than 1 Gyr, 15.93\% have ages
between 1 and 5 Gyr, and the oldest stars with ages larger than 5 Gyr
are 82.82\%.  Different prescriptions for the stellar velocities would
increase (smaller velocities) or decrease (larger velocities) the
contribution of the old stars to the G-dwarf metallicity distribution
function. The main effect of the stellar migration, in the way in
which it is implemented here, is to increase the percentage of old
stars in the solar neighbourhood compared to the model
without stellar migration.

 Finally, we test the effects on the metallicity distribution of long-lived stars in the solar neighbourhood of `extreme' stellar migrations, where we do not allow stars born in situ at 8 kpc to migrate toward other regions. We distinguish 3 cases:
\begin{itemize}
\item case 1): 10\% of the stars  born at 4 kpc and 20 \%  of those born at 6 kpc migrate towards 8 kpc;
\item case 2): 20\% of the stars born at 4 kpc and 40 \%  of those born at 6 kpc migrate towards 8 kpc;
\item case 3): all the stars  born at 4 and 6 kpc migrate towards 8 kpc. 
\end{itemize}

We show (Figure \ref{ex_gd}) that the main effect of stellar migration
in all these extreme cases (even in the really extreme and unphysical
case 3) is to populate the high-metallicity tail of the predicted
distribution. Overall, the predicted stellar metallicity distribution
is not substantially affected. In particular, the position of the peak
is left unchanged: it mostly depends on the assumed timescale of
thin-disc formation. We conclude that the G-dwarf metallicity
distribution cannot be used to extract information on the extent and
significance of stellar migration. However, this process is clearly
needed in order to improve the agreement between model predictions and
observations on the high-metallicity tail.

\section{Conclusions}

In this paper we have studied the effects of the stellar migration by
means of a detailed chemical evolution model where   radial gas flows
in the galactic disc are also considered. The stellar migration is
implemented in the code in a phenomenological way, just mimicking the
results of  published dynamical models,  with no intention of studying the physics of this phenomenon. Our main conclusions can
be summarized as follows:

\begin{itemize}
\item Our reference chemical evolution model for the thin disc of the Milky Way,
assuming a variable star formation efficiency, a constant radial gas inflow 
(fixed at -1 km s$^{-1}$) and no stellar migration, is able
to well fit the majority of the observables: the abundance gradient
along the Galactic disc, the stellar mass surface density profile; also
the observed G-dwarf metallicity distribution of thin-disc stars 
(Adibekyan et al. 2013) is generally in agreement with our model, the only
problem residing in the high-metallicity tail of the 
distribution, which our model is underestimating.

\item By including stellar migration according to simple prescriptions 
that mimic the results of chemo-dynamical models by
Minchev et. al. (2013), and taking into account velocities of 1
km s$^{-1}$ for the migrating stars, we are able to reproduce very 
well the high-metallicity
tail of the G-dwarf metallicity distribution observed for thin-disc stars.

\item By exploring the velocity space, we conclude that the stellar
migration velocities must have values larger than 0.5 km s$^{-1}$
and smaller than 2 km s$^{-1}$ to properly reproduce the high-metallicity 
tail in the G-dwarf metallicity distribution of thin-disc stars.

\item As found by previous authors (Fran\c{c}ois \& Matteucci
1993; Sch\"onrich \& Binney 2009a; Minchev et al. 2013) the
observed spread, if real, in [O/Fe] vs [Fe/H], and age vs [Fe/H] relations,
can be partially explained by taking stellar
migration into account. We proved that the hypothesis of
stellar migration as a solution for the observed spread requires
necessarily a variable SFE.
\end{itemize}

\section*{Acknowledgments}
We thank the anonymous referee for his/her suggestions that have improved the paper. We acknowledge financial support from PRIN MIUR 2010-2011, project `The Chemical and Dynamical
Evolution of the Milky Way and Local Group Galaxies', prot.~2010LY5N2T.

\end{document}